**Title:** "SHIFT AND BROADENING OF SODIUM *n*S–3P AND *m*D–3P TRANSITIONS IN HIGH PRESSURE NaCd AND NaHg DISCHARGES"

**Authors:** ZELJKA MIOKOVIC[*], DARKO BALKOVIC and DAMIR VEZA[#]

**Address:** Physics Department, Faculty of Science, University of Zagreb, Bijenicka 32, HR-10002 Zagreb, Croatia

**Short title:** "Shift of sodium *n*S–3P and *m*D–3P transitions"

**PACS:** 52.70.Kz, 52.80.Mg, 52.25.Os

**Abstract.** We report measurements of the Stark-shift and Stark-width of sodium spectral lines corresponding to $n^2S_{1/2} - 3^2P_{3/2;1/2}$ $(n = 5,6,7)$ and $m^2D_{5/2;3/2} - 3^2P_{3/2;1/2}$ $(m = 5,6)$ transitions. Measurements have been made on AC driven high pressure Na-Hg-Xe and Na-Cd-Xe discharges. Electron density has been determined from the measurements of the Stark-shift of sodium $7^2S_{1/2} - 3^2P_{1/2;3/2}$ spectral line. The electron temperature and the density of neutral particles have been determined analyzing the shape of the measured sodium atomic lines, by comparison of the calculated and the observed line profiles. Electron density and temperature range from 6×10²¹ to 1×10²² m⁻³ and from 3800 to 4100 K, respectively. Dependences of measured Stark-shifts and Stark–widths on electron density and temperature have been investigated. The final results have been compared with available theoretical and experimental data.

---

[*] permanent address:
Faculty of Electrical Engineering, University of Osijek, K. Trpimira 2B, 31000 Osijek, Croatia

[#] corresponding author, e-mail: veza@phy.hr

## 1. Introduction

The high-pressure sodium-mercury vapor discharge lamp was invented 40 years ago [1], and further developed and refined during the next two decades [2]. It is a wall-stabilized electric arc burning at elevated pressure and temperature in mercury-sodium plasma. The lamp burner is made of a high–density polycrystalline alumina ceramics (PCA), a material highly resistant to hot and corrosive sodium vapor up to 1800K. Despite of its high transmission coefficient (> 90% for visible radiation), a disadvantage of PCA is that it is not transparent but translucent material. Consequently, conventional plasma diagnostic techniques for determination of various plasma parameters are not easily applicable to this radiation source.

Optical and electrical features of the high-pressure sodium discharge lamp based on the standard sodium-mercury-xenon filling, have been widely investigated [2]. A similar discharge, based on sodium-cadmium-xenon filling, has been studied so far in order to find out origin of continuous bands in visible part of the spectrum of NaHg, KHg and NaCd molecules [3, 4], but plasma parameters of NaCd discharge have been rarely investigated [5]. Yet, these data are desirable for a more complete understanding of the physics and chemistry of all high-pressure discharges containing sodium and IIB group elements (Zn, Cd, Hg). This general knowledge about plasma generated in a vapor mixture of sodium and IIB group elements is important for a possible (ecological) upgrade to a high-pressure mercury-free discharge lamp based on the sodium-zinc filling [6]. Attractive candidates for determination of electron density and temperature in such plasmas are higher-lying sodium excited states (belonging to *nS* and *mD* series) with large Stark-shifts and Stark-widths and small oscillator strengths. Theoretical Stark-broadening data on sodium $m^2D - 3^2P$ and $n^2S - 3^2P$ transitions are given by Griem's [7, 8] and Dimitrijevic [9] calculations. Experimental data are rather scarce. By our knowledge there is an older measurement of electron density in Na-Hg-Xe discharge using Stark-shifts of $5^2S - 3^2P$, $4^2D - 3^2P$ and $5^2D - 3^2P$ transitions [10]. Just recently the new measurements of the shifts of $7^2S - 3^2P$, $5^2D - 3^2P$ and $6^2D - 3^2P$ spectral lines in a Na-Ar pulsed discharge are published [11]. The accompanied calculations of these line-shifts [11] made in the random-phase approximation (RPA) agree with existing Stark-shift calculations [7, 8, 9]. Mutual agreement between theories [9] and [11], and experiment [11] is especially good in the case of the $7^2S - 3^2P$ transition, suggesting that this line is a very good choice for electron density measurements in all discharges containing sodium as an additive.

In this paper we report the measurements of the Stark-shift and Stark-width of sodium spectral lines corresponding to $n^2S - 3^2P$ $(n = 5,6,7)$ and $m^2D - 3^2P$ $(m = 5,6)$ transitions, originating in NaCd and NaHg high-pressure discharges. Because of very good agreement between theory and experiment in the case of the $7^2S_{1/2} - 3^2P_{1/2;3/2}$ line [11], we used this line as a calibrant line, i.e. the electron density has been determined from the measurements of the Stark-shift of this spectral line. We investigated the dependence of Stark-shift and Stark-width of the sodium $n\text{S}-3\text{P}$ $(n=5,6)$ and $n\text{D}-3\text{P}$ $(n=5,6)$ atomic lines on electron density. The measured data are compared with the available experimental [10, 11] and theoretical data [7, 8, 9, 11]. The observed line shapes are compared with the calculated line shapes obtained using Bartels' method [12].

## 2. Line-shape calculations

2.1. Broadening mechanisms

The spectral line profiles of sodium nS-3P and mD-3P transitions radiated by excited sodium atoms in a high pressure arc discharge are influenced by two main broadening mechanisms: the Stark broadening (by electrons) and the van der Waals broadening (by dissimilar atoms) [7, 13]. The resonance broadening (by similar atoms) and the Doppler broadening are practically negligible under our experimental conditions [7, 14]. However, the dominant shift and broadening mechanism in this experiment is the Stark broadening by electrons, as it will be justified later.

The shape of an atomic line can be calculated in two different approximations, impact and quasistatic, corresponding to two different interaction pictures. However, the application of the impact or the quasistatic approximations yields very different line shapes [7]. The *quasistatic approximation* yields the following line profile [14]:

[1]
$$P_{\text{QS}}(\Delta\lambda) = \frac{\sqrt{\Delta\lambda_0}}{2\,\Delta\lambda^{3/2}}\,\exp\!\left(-\frac{\pi\,\Delta\lambda_0}{4\,\Delta\lambda}\right),$$

where $\Delta\lambda > 0$, and $\Delta\lambda_0$ represents the characteristic width of this profile. This profile yields a broadening of the red wing of sodium lines, and there is no contribution in the blue line-wing: $P_{\text{QS}}(\Delta\lambda \leq 0) = 0$. The characteristic width of the quasistatic profile, $\Delta\lambda_0$, and the shift (displacement of the maximum), $\delta\lambda_{\max}$, due to collisions of the radiating sodium atom with neutral dissimilar atoms (the van der Waals interaction) are given as [14, 8]:

[2]
$$\Delta\lambda_0 = \frac{\lambda_0^2}{2\pi c}\left(\frac{4\pi}{3}\right)^2 C_6^{\text{NaX}}\, N_{\text{X}}^2,$$

[3]
$$\delta\lambda_{\max} = \frac{\lambda_0^2}{2\pi c}\frac{\pi}{6}\left(\frac{4\pi}{3}\right)^2 C_6^{\text{NaX}}\, N_{\text{X}}^2.$$

Here $\lambda_0$ represents the transition wavelength, $C_6^{\text{NaX}}$ represents the effective Na-X interaction constant, and X denotes cadmium or mercury atom. The effective $C_6^{NaX}$ interaction constants depend linearly on the cadmium (mercury) dipole polarizability $\alpha_X$ [15, 16] as $C_6^{NaX} = \alpha_X \, e^2 \left(\langle r_i^2 \rangle - \langle r_f^2 \rangle\right)$, where $\langle r_i^2 \rangle$ and $\langle r_f^2 \rangle$ are the mean square radii of the valence electron in the initial and the final excited sodium state [14].

On the other hand, the *impact approximation* yields the Lorentzian line shape regardless the broadening mechanism [7]:

$$[4] \qquad P_L(\Delta\lambda) = \frac{\Delta\lambda_{1/2}}{2\pi} \frac{1}{\left(\Delta\lambda - \delta\lambda_{1/2}\right)^2 + \left(\Delta\lambda_{1/2}/2\right)^2},$$

where $\delta\lambda_{1/2}$ is the line shift, and $\Delta\lambda_{1/2}$ is the full half-width (FWHM) of the Lorentz profile. The line-width and line-shift of the Lorentzian profile in our case are due to Stark and van der Waals interactions [2, 7]. The total line-width is given by the sum of the Stark and van der Waals line widths, $\Delta\lambda_{1/2} = (\Delta\lambda_{1/2})_S + (\Delta\lambda_{1/2})_{vdW}$. The total line-shift is given by $\delta\lambda_{1/2} = (\delta\lambda_{1/2})_S + (\delta\lambda_{1/2})_{vdW}$ [7, 8, 14].

According to the Lindholm-Foley impact theory [14] the Stark broadened line full half-width, $(\Delta\lambda_{1/2})_S$, and the line shift, $(\delta\lambda_{1/2})_S$, are given by:

$$[5] \qquad (\Delta\lambda_{1/2})_S = \frac{\lambda_0^2}{2\pi c} 11.37 \, C_4^{2/3} \, v_e^{1/3} \, N_e,$$

$$[6] \qquad (\delta\lambda_{1/2})_S = \frac{\lambda_0^2}{2\pi c} 9.94 \, C_4^{2/3} \, v_e^{1/3} \, N_e.$$

The $C_4$ represents the Stark interaction constant, $v_e$ the mean electron velocity, and $N_e$ is the electron density. More sophisticated, quantum-mechanical calculations [7] lead to the following dependence of the Stark broadened line full half-width, $(\Delta\lambda_{1/2})_S$, and the line shift, $(\delta\lambda_{1/2})_S$, on electron density:

$$[7] \qquad (\Delta\lambda_{1/2})_S = 2\left[1 + 1.75 A(1 - 0.75 R)\right] w_e N_e,$$

$$[8] \qquad (\delta\lambda_{1/2})_S = \left[\frac{d_e}{w_e} \pm 2 A(1 - 0.75 R)\right] w_e N_e.$$

The Stark half-halfwidth and -shift data due to electron broadening ($w_e$ and $d_e$ values), as well as the ion broadening parameter $A$, are tabulated in [7], [8] and [9] for a reference electron density of $10^{22}$ m$^{-3}$ and a broad temperature range. Parameter *R* representing the ratio of the mean distance between ions to the Debye radius [7] is given by:

[9]
$$R = \frac{r_1}{\rho_D} = 6^{1/3}\pi^{1/6}\left[\frac{e^2}{4\pi\varepsilon_0 kT_e}\right]^{1/2} N_e^{1/6}.$$

According to the Lindholm-Foley impact theory [14], a line broadened by van der Waals interaction has the full half-width, $(\Delta\lambda_{1/2})_{\text{vdW}}$, and the line shift, $(\delta\lambda_{1/2})_{\text{vdW}}$, given by:

[10]
$$(\Delta\lambda_{1/2})_{\text{vdW}} = \frac{\lambda_0^2}{2\pi c} 8.08 \left(C_6^{\text{NaX}}\right)^{2/5} v^{3/5} N_X \text{ , and}$$

[11]
$$(\delta\lambda_{1/2})_{\text{vdW}} = \frac{\lambda_0^2}{2\pi c} 2.94 \left(C_6^{\text{NaX}}\right)^{2/5} v^{3/5} N_X.$$

Here $v$ is the relative velocity of interacting atoms, $C_6^{\text{NaX}}$ represents the effective Na-X interaction constant, and X stands for Cd or Hg. The $C_6^{\text{NaX}}$ interaction constants are calculated using the data given in [15, 16, 17].

According to Kielkopf [18] and Stormberg [13], the total line profile can be successfully simulated by convoluting the impact line profile and the quasistatic line profile:

[12]
$$P_T(\Delta\lambda) = \int P_L(\Delta\lambda - \varsigma) P_{QS}(\varsigma) d\varsigma.$$

The convolution integral gives a line profile expressed in a complicated but analytical form that can be rapidly calculated [19]. The resultant synthetic line profile describes simultaneously the line core and the line wings. This empirical approach has been suggested long time ago by Margenau [20], Hindmarsh et al [21], Kielkopf et al [18], and recently by Stormberg [13]. However, whereas this approach is mathematically correct, it can be questioned from the physical point of view. It suggests that we can convolute two broadening mechanisms that are rigorously limited: one to the line core (giving the impact line profile), and the other one to the line wing (delivering the quasistatic line profile). This is certainly not true [8], but justification for the application of this synthetic profile and its popularity in line shape analysis [22-24], comes from the fact that it smoothly and correctly describes the complete line shape, behaving as a Lorentzian in the line core and in the blue line wing, and as an exponential in the red line wing. After Stormberg [13] several authors successfully applied this approach to the analysis of the shape of atomic lines radiated by various high-pressure discharges (sodium lines in high-pressure sodium and metal-halide discharges [16, 22, 23], and mercury lines in high-pressure mercury discharge [24]). In this work the line profiles were calculated by the line shape function $P_T(\Delta\lambda)$ and the synthetic line shape was used in Bartels' method to obtain the true line shape emitted by the discharge.

## 2.2. Bartels' method

This method [25], is based on the assumptions that the axially symmetric plasma is in local thermodynamic equilibrium (LTE), the equation of state for ideal gas fulfilled, the partial pressure of the emitting atoms constant throughout the plasma column, and that the depletion of the ground state population due to excitation and ionisation can be neglected.

The population of an excited atomic state is then given by the Boltzmann equation:

$$[13] \quad N_n(r) = N_0(r) \frac{g_n}{g_0} \exp\left(-\frac{E_n}{kT(r)}\right),$$

where $g_n$ and $g_0$ are statistical weights of level $n$ and of the ground state, respectively. The ground–state atomic density, $N_0(r)$ is related to the total vapour pressure and the gas temperature via the equation of state

$$[14] \quad N_0(r) = \frac{g_0}{U(T)} \frac{p}{kT(r)},$$

where $U(T)$ is the partition function of the atom, $k$ is the Boltzmann constant and $p$ is the local vapor pressure. Under high-pressure discharge conditions the partition function $U(T)$ is approximately equal to $g_0$. $T(r)$ is the temperature distribution along a radius of the discharge

$$[15] \quad T(r) = T_A - (T_A - T_W)\left(\frac{r}{R}\right)^n,$$

where $T_A$ is the axis temperature, $T_W$ is the wall temperature, and $n \cong 2$ (a parabolic temperature distribution).

Finally, the intensity of a spectral line within Bartels' method is given by:

$$[16] \quad I(\nu) = \frac{2h\nu^3}{c^2} \exp\left(-\frac{h\nu}{kT_m}\right) M\, Y(\tau_0, p),$$

where $\tau_0$ is the optical depth, and the function $Y(\tau_0, p)$ represents the influence of the optical depth on the peak line intensity (could be expressed parametrically). The parameters $M$ and $p$ describe the inhomogeneity of the plasmas, so that $p = 1$ corresponds to a homogeneous plasma column, and $p = 0$ to a completely inhomogeneous source. The energy of the lower level for the line under investigation must satisfy the conditions: $kT \ll E_n$ (pressure broadening case) and $kT \ll E_n + 0.5E_i$ (Stark broadening case). The function $M$ must satisfy conditions: $M = \sqrt{E_n/E_m}$ (van der Waals broadening) and $M = \sqrt{E_n + 0.5E_i}/\sqrt{E_m + 0.5E_i}$ (Stark broadening). $E_n$ and $E_m$ are the energies of the lower and the upper level of the atomic line, respectively, whereas $E_i$ is ionisation energy. Both, $M$ and $p$, are constant within a spectral line.

## 3. Experimental setup

The experimental arrangement is shown in Fig. 1. The measurements have been performed using 400 W NaCd and NaHg high pressure discharges (with translucent sapphire burner) in series with a 400 W inductive choke. The inner diameter, the outer diameter and the length of the sapphire burner are 7.6 mm, 9 mm and 110 mm, respectively. The tips of the electrodes are separated 95 mm. The discharge was operated vertically, driven by a standard 50 Hz AC line source, with a discharge current between 3 A and 4.2 A for the NaCd , and between 2.6 A and 4A for the NaHg high-pressure lamp. The burner contains a sodium–cadmium (or sodium-mercury) amalgam. Partial pressure of sodium and cadmium (or mercury) during lamp operation is determined by the temperature of the coldest spot of the burner [26]. The central region of the burner was imaged onto the entrance slit of a medium-resolution monochromator ("K. Zeiss", model SPM-2, 10 μm wide slits). The magnification ratio was approximately 1:1. The light from the discharge, spectrally resolved by the monochromator, was detected using an EMI 9558QB photomultiplier. A linear amplifier (Keithley electrometer, M611) with a low-pass electronic filter for the suppression of high-frequency noise has been used for signal amplification.

Signal is processed by a box-car averager equipped with a gated integrator, operated either in external- or in line-triggering mode as illustrated in Fig. 2. The current reversal point is the reference time instant and the origin of the trigger. The trigger signal can be derived either by monitoring the integral light signal by a photodiode, or by monitoring the AC current. This technique enables sampling out the spectrum of plasma radiation at any desirable phase of the AC current. In this experiment we performed all the measurements using the aperture duration time of 50 μs, and the aperture delay time of 5 ms (at the current maximum). The processed data have been converted by a 14-bit A/D converter, and stored in a computer for further analysis.

Calibration of the monochromator wavelength scale is provided by low-pressure sodium and mercury lamps. The shift of spectral lines radiated by the low-pressure plasma can be neglected in this case.

## 4. Plasma diagnostics and reduction of data

To enable comparison of our data with the results of earlier experiments [10, 11] and with calculations of Stark-widths and Stark-shifts made under assumption of LTE [7, 8, 9, 11] we have to check validity of LTE approximation in our AC driven arc discharge. Here we will show that our

Miokovic et al

discharge satisfies all validity criteria for assumption of LTE, and that the data may be compared to data measured in previous experiments.

In the experiment [10] authors measured Stark-widths and Stark-shifts on $5^2S - 3^2P$ and $5^2D - 3^2P$ emission lines from a NaHg arc burning in a tube made of clear sapphire (electrode separation 8.6 cm, internal diameter 7.6 mm). The arc was operated with a switching DC power source, with 80 ms switching time. The measured electron densities in the core of the arc were about 1x10$^{21}$ m$^{-3}$, justifying the assumption of LTE in their measurements. In the recent experiment [11] authors measured Stark-shifts on $7^2S - 3^2P$, $5^2D - 3^2P$ and $6^2D - 3^2P$ lines from a Na-Ar arc burning in a clear sapphire tube (electrode separation 7.6 cm, internal diameter 4.8 mm). The arc was operated in a simmer-mode, powered with rectangular current pulses of 1 ms duration [11]. Electron densities in the core of the arc were measured as high as 1x10$^{23}$ m$^{-3}$, justifying the assumption of LTE in their measurements, too. The electron density range, the core arc temperature and the density of neutral atoms in these discharges are very similar to our experimental conditions.

A comprehensive discussion of validity criteria for assumption of LTE can be found in [7]. In case of stationary plasma, for complete LTE down to the ground state, the collision excitation rate must be much larger than the radiative excitation or de-excitation rate even for the first excited state. Since the resonance lines of the main discharge constituent (neutral mercury or cadmium) are self-absorbed, the condition for validity of LTE requires ([7], Eq. 6-60):

[17] $$N_e \geq 9*10^{17} \sqrt{\frac{kT}{E_H}} \left(\frac{E_2 - E_1}{E_H}\right)^3,$$

where the symbols have the same meaning as in the reference. This condition requires rather high electron densities, but can be reduced for an order of magnitude if the product of ground state atom density, $N_1$, and the discharge diameter, $d$, is high enough ([7], Eq. 6-64):

[18] $$N_1 * d \geq 4*10^9 f_{12}^{-1} \lambda_{12}^{-1} \sqrt{\frac{kT}{A\, E_H}},$$

where $f_{12}$ denotes the resonance line oscillator strength, $\lambda_{12}$ is the wavelength of the line, and $A$ represents the relative atomic mass number. Since the Eq. [18] is satisfied in our experiments, the Eq. [17] may be safely relaxed for an order of magnitude:

[19] $$N_e \geq 10^{17} \sqrt{\frac{kT}{E_H}} \left(\frac{E_2 - E_1}{E_H}\right)^3,$$

which for neutral mercury or cadmium (the most stringent requirement) becomes $N_e > 7*10^{20}$ m$^{-3}$. This basic requirement for validity of LTE is always satisfied in our measurements since in all experiments the line widths and shifts are determined at $N_e \geq 1*10^{21}$ m$^{-3}$. However, to prove validity of LTE one must additionally check whether the kinetic temperatures of atoms and ions are equal to the electron temperature. This condition is given as ([7], Eq. 6-75):

[20] $$E^2 << \left(5.5*10^{-12} N_e \frac{E_H}{kT}\right)^2 \frac{m}{M},$$

where $E$ is the applied electric field, and the other symbols have the same meaning as in the reference. Since this condition is satisfied we can expect that the temperatures of heavy particles are equal to electron temperature to within ±2% [7].

Since our discharge is AC operated, and data sampled at each current maximum point with 50 μs aperture duration (see Fig. 2.) we also have to check the validity of LTE in the case of time-dependent plasmas. Theory requires that the time necessary for equilibration between atomic states be less then a typical time for a 10% variation in plasma conditions during that period in which we make the measurement. The time required for equilibration between states with quantum numbers $n$ and $n+1$ is given as ([7], Eq. 6-67):

[21] $$t \geq t_{crit}; \quad t_{crit} = \frac{4.5 \times 10^7}{n^4 N_e} \sqrt{\frac{kT}{E_H}} \exp\left(\frac{E_{n+1} - E_n}{E_H}\right),$$

where the symbols have the meaning as in the reference [7]. For the ground state and the first excited state of neutral mercury (or cadmium), what is again the most stringent requirement, this formula yields $t_{crit}$ = 60 μs. Since typical time allowed for plasma variation in our experiment is 50μs (aperture duration time) this condition is barely satisfied for resonance levels of neutral mercury (or cadmium), but it is entirely satisfied for all higher atomic states of all constituents of our plasmas.

Finally, we have to check conditions of validity of the isolated line approximation [7] for the sodium lines of interest. Generally, at lower electron densities the line shifts due to quadratic Stark effect are proportional to electron density. Experiments in dense plasmas, made at higher electron densities, show deviations from this linear behaviour. The determination of the electron density at which the deviation from linearity can start is very important since it establishes range of validity of the isolated line approximation. By definition, [7, 9], a line can be regarded as isolated if its initial and final (non-degenerate) energy levels $E_i$ and $E_f$ broadened by electron collisions do not overlap, i.e. if $2w^i_e \leq \omega_{in}$ ($2w^f_e \leq \omega_{fm}$). Here $2w^i_e$ and $2w^f_e$ are corresponding level widths, and $\omega_{in}$ (or $\omega_{fm}$) is the distance to the corresponding nearest perturbing level, $E_n$ and $E_m$. If a line is isolated the condition $w_e \leq \omega_{jj'}$ must be satisfied. Here $w_e$ represents the line-width and $\omega_{jj'}$ is the energy distance to the nearest perturbing level. To check if a line is isolated we can use the parameter $C$ defined as $C = 2w_e/(\Delta E_{ij})_{min}$ and listed in [9] for each sodium line. The "critical" electron density $N_l$, can be expressed as $N_l = C/2w_e$, where $w_e$ is the line half-width at the reference electron density of $1*10^{22}$ m$^{-3}$. It can be easily checked that for all nS-3P lines departure from linear behavior can be expected at very high electron densities ( $N_l \geq 1*10^{23} m^{-3}$ ). This suggests that all nS-3P lines can be safely treated as isolated in the electron density range in our experiments. On the contrary, the value of $N_l$ for the 5D-3P line it is only about 1*10$^{22}$ m$^{-3}$. For the 6D-3P line it is even lower, 4*10$^{21}$ m$^{-3}$, suggesting that these two lines could not be regarded as completely isolated.

In the case of $n^2S - 3^2P$ $(n = 5,6,7)$ and $m^2D - 3^2P$ $(m = 5,6)$ sodium lines, and in the discharge parameter range of interest (3800 K ≤ $T_e$ ≤ 4100 K, 6*10$^{21}$ < $N_e$ < 1*10$^{22}$ m$^{-3}$, 1*10$^{23}$ < $N_{Na}$ < 4*10$^{23}$ m$^{-3}$, 5*10$^{23}$ < $N_X$ < 2.5*10$^{24}$ m$^{-3}$, where "X" stands for Cd and Hg) the Stark broadening dominates over van der Waals broadening. That statement can be proved by inserting appropriate interaction constants, plasma parameters and other experimental data in the equations [2] – [11]. To demonstrate it, we can use the data on the 6S-3P line measured in NaHg and NaCd discharges at an electron density of about 9*10$^{15}$ cm$^{-3}$. The corresponding discharge temperature is determined to be 3800 K in NaHg and 3900 K in NaCd discharge. The mean relative velocity of Na-Hg (or Na-Cd) atomic pair at this temperature is about 2*10$^3$ m/s. An average density of ground state mercury (or cadmium) atoms at this temperature is about 2*10$^{24}$ m$^{-3}$, and an average ground state sodium atom density is about 3*10$^{23}$ m$^{-3}$. The static polarizability of mercury atoms is 34.4 a.u., and 49.7 a.u. for cadmium atoms [15, 16]. The corresponding effective C$_6$ constant for Na$^*$-Hg pair is about 4*10$^{-29}$ s$^{-1}$ cm$^6$, and for Na$^*$-Cd pair is about 5*10$^{-29}$ s$^{-1}$ cm$^6$. These values lead to an impact line-shift (caused by Cd or Hg perturbers) of about 0.006 nm , and to a displacement (also caused by Cd or Hg perturbers) of the quasistatic profile of about 0.00025 nm. The corresponding typical line-broadening values are about 0.017 nm for the impact line-width, and about 0.0005 nm for the characteristic width of the quasistatic profile. Furthermore, the Na(6S) - Na(3P) interaction, according to the Lindholm-Foley impact theory [14], with a calculated interaction constant of about 2*10$^{-10}$ s$^{-1}$ cm$^3$, delivers a completely negligible line-width, and a null line-shift. Also, the Doppler broadening at this temperature contributes about 0.004 nm to the line-width, and also causes a null line-shift.

At indicated electron density (N$_e$ = 9*10$^{15}$ cm$^{-3}$) the measured $6^2S - 3^2P$ total line-shifts are about 0.08 nm. Using plasma parameters indicated above, our fit procedure delivers a Stark-widths for this transition of about 0.045 nm. The analogous calculated shift and broadening data for other two $n^2S - 3^2P$ and $m^2D - 3^2P$ lines are basically very similar. Consequently, the line-shifts of all $n^2S - 3^2P$ and $m^2D - 3^2P$ transitions are really heavily dominated by Stark shift: $(\delta\lambda_{1/2})_S \gg (\delta\lambda_{1/2})_{vdW}$ . The corresponding line-widths are significantly dominated by Stark broadening: $(\Delta\lambda_{1/2})_S > (\Delta\lambda_{1/2})_{vdW}$ . Furthermore, since typical ratio of the mean distance between ions and the Debye radius is $R \leq 0.65$ (Eq. [9]), the contribution of ion broadening to the total line-width and -shift of $n^2S - 3^2P$ lines should be less than 10%. Therefore, one can assume for the Stark broadening and shift parameters (Eqs. [7] - [8]) a linear dependence on electron density. On the other hand, the contribution of ion broadening to the total line width and shift of the $n^2D - 3^2P$ lines could be up to 25%, and one cannot assume for Stark broadening parameters a safe linear dependence on electron density.

The principle of the measurement of line-shifts is shown in Fig. 3 [26]. The line-positions measured from the high-pressure sodium discharge and from the low-pressure sodium lamp can be directly compared, delivering Stark shift, $(\delta\lambda_{1/2})_S$, of the corresponding line profiles. On the contrary, the Stark widths, the discharge temperature and the density of neutrals are determined indirectly, by fitting the synthetic line shapes to the measured line profiles as shown in the Fig. 4. The 6S-3P line shown in this figure is measured in NaCd discharge at an electron density of $(9\pm2.3)*10^{21}$ m$^{-3}$. The discharge temperature is determined to be 3900 ± 100 K. The mean relative velocity of Na-Cd pair at this temperature is $2.07*10^3$ m/s. The density of ground state cadmium atoms is $(1.8\pm0.6)*10^{24}$ m$^{-3}$, and the ground state sodium atom density is $(2.7\pm0.8)*10^{23}$ m$^{-3}$. The static polarizability, α, of cadmium atoms is 49.7 a.u. [15]. The corresponding effective $C_6$ constant for Na$^*$(6S)-Cd pair is $5.4*10^{-29}$ s$^{-1}$ cm$^6$. These values lead to an impact van der Waals line-shift of 0.0055 nm, and to a displacement of the quasistatic profile of 0.00022 nm. Corresponding line-broadening values are 0.015 nm for the impact line-width, and 0.00043 nm for the characteristic width of the quasistatic profile. At electron density of $9*10^{21}$ m$^{-3}$ the $6^2S - 3^2P$ total line-shift is measured as (0.08±0.02) nm, and our fit procedure delivers (0.044±0.011) nm as the Stark-width of this transition. Since the line-shift of $6^2S - 3^2P$ transition is completely dominated by Stark shift, $(\delta\lambda_{1/2})_S \approx 15(\delta\lambda_{1/2})_{vdW}$, the measured total line shift practically corresponds to Stark shift.

In addition, we would like to note that synthetic line shapes calculated within the Bartels' method were numerically convoluted [27] with the (0.13 ±0.01) nm wide Gaussian instrumental profile (determined in a separate experiment) in order to obtain the profile that can be directly compared to the measured one. The full line in Fig. 4 represents the best fit. The dashed (dotted) line represents the fit with 25% larger (smaller) Stark-width. The small picture shows the difference between corresponding calculated line shapes and experimental line-shape. Note that the best fit profile fits almost exactly the blue wing of the weaker component, and also very well the region of the blue wing of the stronger component. Both components show a slight residual discrepancy between calculated and measured red line-wing shape (maximum 8% difference in the red wing). We strongly believe that this residual line-shape discrepancy could be caused by quasistatic ion broadening.

## 5. Results and Discussion

In Figs. 5a-c results of our measurements of line shifts of nS-3P lines (n=5,6,7) are compared with available theoretical data [7, 8, 9]. Note that electron densities in the NaCd and NaHg discharges are derived from the shift data of the 7S-3P line (Fig. 5a) with the Stark broadening parameters calculated by [9] and [11]. Traditionally, the line shift data are regarded as less reliable for determination of electron density in a discharge because of the difficulties in making accurate theoretical predictions. Since several effects with contributions of different sign could be responsible for the Stark shift of spectral lines, the theoretical shift parameters usually have a larger uncertainty. Fortunately, in the case of the 7S-3P transition two main contributions to the line shift, stem form the interaction with the 7p and 6p levels, and lead to the shifts of opposite sign [11]. Their energy difference to 7s level are about –0.085 eV and +0.062 eV, respectively. In addition, their sum is almost linear with respect to electron density. The random-phase (RPA) calculations of the 7S-3P line-shift [11] completely confirm the line-shift data of Dimitrijevic [9]. Apparently, the 7S-3P line is well suited to measure electron density in high-pressure discharges containing sodium in the range of electron densities up to $10^{23}$ m$^{-3}$. Consequently, our experimental points of the measured shift data for the 7S – 3P line are placed exactly on the line given by theoretical calculations of Dimitrijevic [9]. However, one can note that both, [7] and [9], calculations are within our experimental error bars (±25%). In addition one should take into account that claimed uncertainty of calculated shift data is about 30% [7]. The 6S – 3P and the 5S- 3P lines show a small but systematic discrepancy between experiment and calculations.

In Figs. 6a-c results of our determinations of line widths of nS-3P lines (n=5,6,7) are compared with the available theoretical data [7, 8, 9]. The experimental points of the Stark-width data for the 7S – 3P line are scattered around the calculations of Dimitrijevic [9]. However, note again that both calculations, [7] and [9], are within our experimental error bars (±25%), and that claimed uncertainty of calculated width data is about 30% [7]. The widths of the 6S – 3P and the 5S- 3P lines show a small but systematic discrepancy between experiment and calculations.

In Figs. 7a-b results of our measurements of line shifts of 6D-3P and 5D-3P lines are compared with the available theoretical [7, 8, 9] and experimental data [11]. In the case of 6D – 3P transition our experimental data are situated closely to [7] and [11] calculations, and far away from the results given in [9]. In this case our data are grouping together with the experimental data given in [11]. In the case of 5D – 3P transition both, [7] and [9] calculations are within our experimental error bars (±25%), but one should also take into account that claimed uncertainty of calculated shift data is about 30% [7]. Apparently our data are grouping along the RPA [11] and [9] theoretical lines. While the estimated contribution of ion broadening to the total line width and shift of $n^2S - 3^2P$ lines is less than our estimated error (±25%), the contribution of ion broadening to the total line width and shift of the $n^2D - 3^2P$ lines could be comparable to our error bars. Furthermore, for all nS-3P lines departure from isolated line approximation can be expected at very high electron densities $N_l \geq 1*10^{23} m^{-3}$, whereas this density for the 5D-3P line it is about $1*10^{22}$ m$^{-3}$. For the 6D-3P line it is even lower, $4*10^{21}$ m$^{-3}$, suggesting that these two lines could not be safely regarded as completely isolated.

Finally, in Fig.8 the measured Stark shifts of sodium nS-3P lines (n=5,6,7) at $N_e=1*10^{22}$ m$^{-3}$ are plotted versus the upper state quantum number together with theoretical Stark shifts derived for the same experimental conditions. There is a smooth dependence of the Stark shifts, both experimental and theoretical, on the upper state quantum number up to quantum number n=7. The experimental data show a systematic disagreement with the theoretical curves towards smaller quantum numbers, but available calculations disagree substantially in the range of higher quantum numbers, starting with n ≥ 8. Additional experiments are needed to clarify noted discrepancies.

## 6. Conclusion

We measured the Stark shifts and widths of neutral sodium spectral lines corresponding to $n^2S - 3^2P$ $(n=5,6,7)$ and $m^2D - 3^2P$ $(m=5,6)$ transitions. The measurements are compared with available theoretical [7, 8, 9, 11] as well as experimental data [11]. The 7S-3P line has been used as a calibrant line to measure the electron density in both, NaHg and NaCd, high-pressure discharges. The data measured for the 6S – 3P and the 5S- 3P lines show a small but systematic discrepancy between experiment and theory. The line shifts of 5D-3P transition are in very good agreement with the available experimental and theoretical data. In the case of 6D – 3P transition our experimental data are in a satisfactory agreement with the existing experimental and theoretical data. The residual discrepancies between experiment and theory can be attributed to the influence of ion broadening and, in the case of the nD-3P transitions, to the breakdown of the isolated line approximation. In addition, the systematic disagreement between theory and experiment towards smaller quantum numbers and substantial disagreement of calculations in the range of higher quantum numbers asks for additional experiments and calculations in order to clarify noted disagreement.

**Acknowledgements**

We gratefully acknowledge the financial support from the Ministry of Science and Technology of the Republic of Croatia (project MZT 0119253), the partial financial support from the European Co-operation in the field of Scientific and Technical Research (COST Project 529) "Light Sources for the 21$^{st}$ Century" and the partial financial support from the joint USA-Croatia project JF107 (NIST)

"Atomic Data for Optimisation of Light Sources". Also, we would like to thank to the reviewers of our paper for a careful reading of the manuscript, for the suggestions they made to improve the quality of the presented article as well as for the new ideas they suggested.

# Figures and Figure captions

**Fig. 1**. The experimental arrangement. HPD – high pressure sodium discharge, LPL – low pressure sodium lamp, F – cut off filter, L – lens, FM – folding mirror, M – monochromator, PMT – photomultiplier, A – linear amplifier, BCA – box car averager, A/D – analog to digital converter, PC – personal computer.

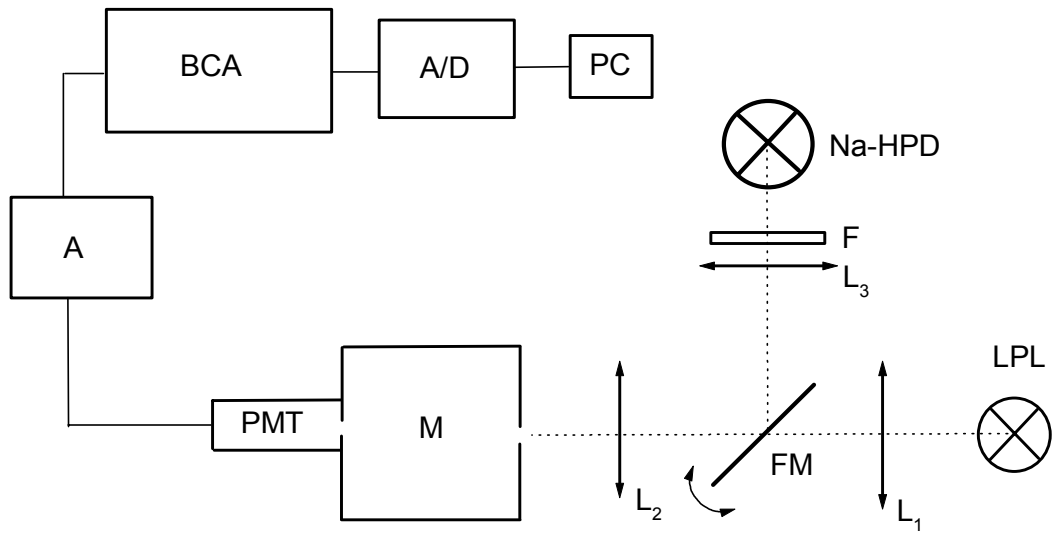

Fig. 1

**Fig. 2.** An illustration of the measurement process. The upper part displays the AC discharge current, the middle part – the corresponding time dependence of the photomultiplier signal, and the lower part – a general case of timing for data acquisition. In our experiments the data are sampled at the current maximum (5 ms aperture delay time) and with 50 μs aperture duration.

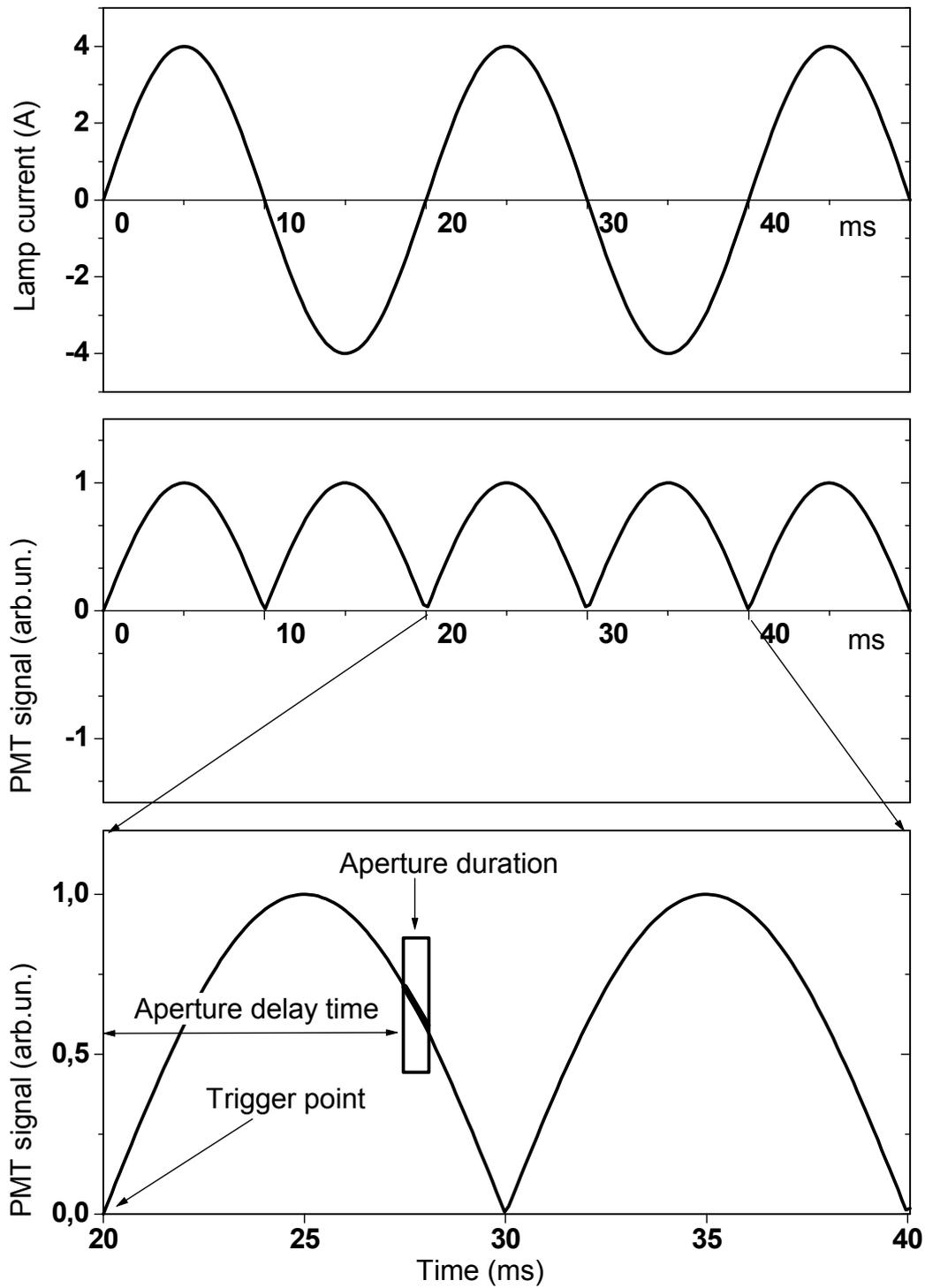

Fig.2

Miokovic et al

**Fig. 3.** Determination of line-shifts. The line-positions measured in the spectrum of high-pressure sodium discharge and in the spectrum of the low-pressure sodium lamp are directly compared, delivering Stark shift $(\delta\lambda_{1/2})_S$, of the corresponding line profiles. In this example the measurement of the total line-shift, $\delta\lambda_e \, B(\delta\lambda_{1/2})_S$, of the 5D - 3P sodium line is illustrated. The small and broad feature (at 497.4 nm) in the blue wing of the weaker component is a forbidden sodium line.

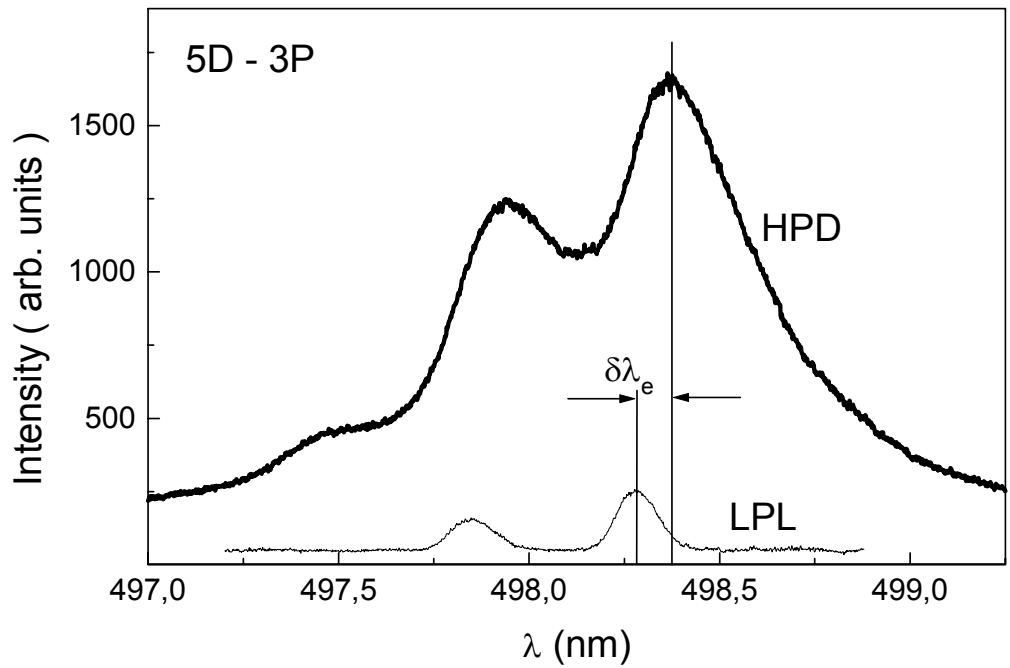

Fig.3

**Fig. 4.** Determination of line-widths. This line is measured in NaCd discharge at $N_e = (9\pm2.3)*10^{21}$ m$^{-3}$. Other discharge parameters are: $T_e = 3900 \pm 100$ K, $N_{Cd}=(1.8\pm0.6)*10^{24}$ m$^{-3}$, $N_{Na}=(2.7\pm0.8)*10^{23}$ m$^{-3}$, effective $C_6^{NaCd} = 5.4*10^{-29}$ s$^{-1}$ cm$^6$. The line-shift of this transition is greatly dominated by Stark shift, $(\delta\lambda_{1/2})_S \approx 15(\delta\lambda_{1/2})_{vdW}$ (see text for discussion). The full line represents the best fit. The dashed (dotted) line represents the fit with 25% larger (smaller) Stark-width. The small picture shows the difference between corresponding calculated line shapes and experimental points.

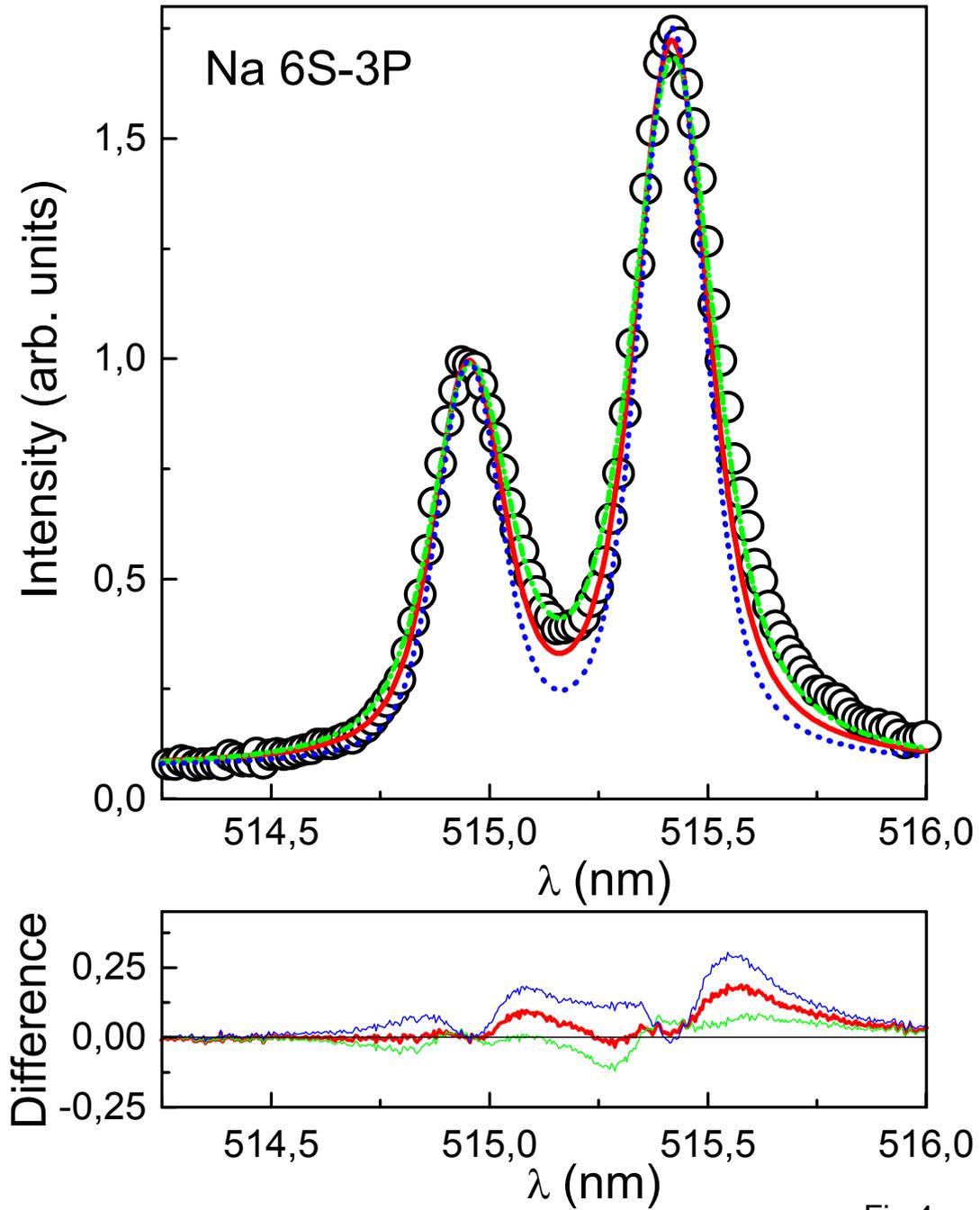

Fig.4

Miokovic et al

**Fig. 5a** Results of the measurements of the Stark-shift for the 7S-3P transition. Explanation of the symbols: ■■■ this experiment, NaHg discharge, ●●● this experiment, NaCd discharge, – – – theory [9], —— theory [7]. This line was used as the calibrant line for further measurements. Estimated experimental error is about ±25%. Claimed uncertainty of calculated shift data is about 30% [7].

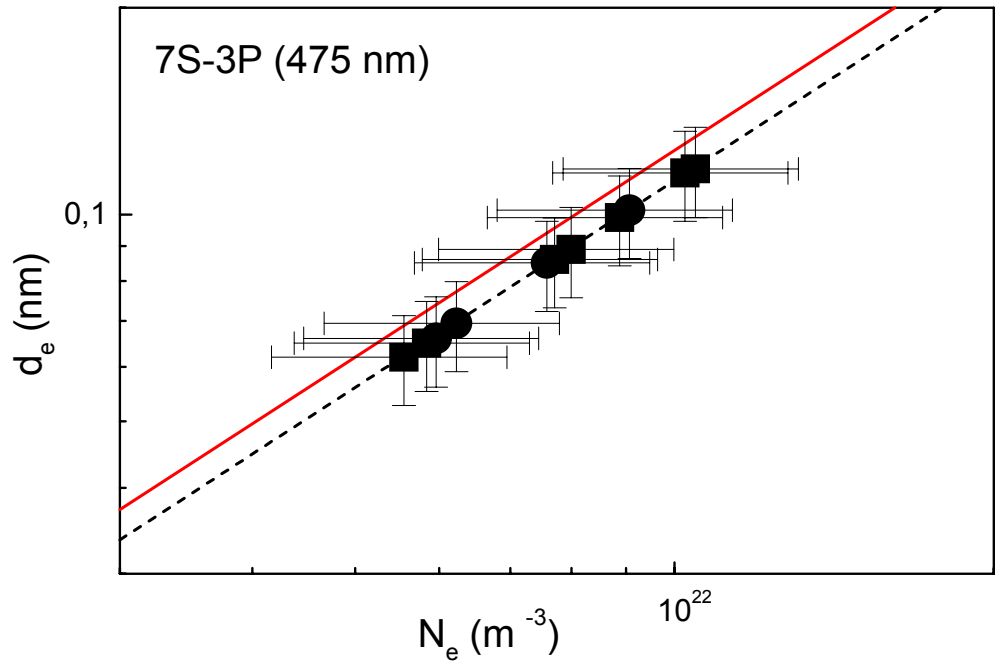

Fig.5a

Miokovic et al

**Fig. 5b** Results of the measurements of the Stark-shift for the 6S-3P transition. Explanation of the symbols: ■■■ this experiment, NaHg discharge, ●●● this experiment, NaCd discharge, – – – theory [9], —— theory [7], –··–··· theory [8]. Note that the calculations of Dimitrijevic [9] and Griem [8] practically coincide. Estimated experimental error is about ±25%. Claimed uncertainty of calculated shift data is about 30% [7].

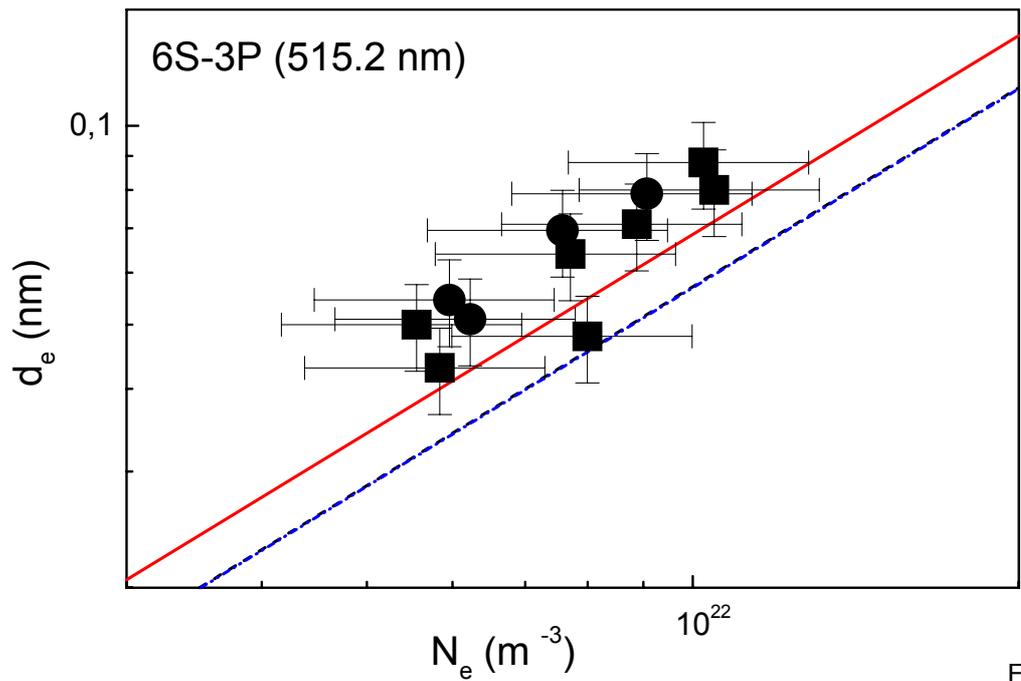

Fig.5b

**Fig. 5c** Results of the measurements of the Stark-shift for the 5S-3P transition. Explanation of the symbols: ■■■ this experiment NaHg discharge, ●●● this experiment NaCd discharge, ▫▫▫ experiment [10], – – – theory [9], —— theory [7], –··–··· theory [8]. Our estimated experimental error is ±25%. Claimed uncertainty of calculated shift data is about 30% [7].

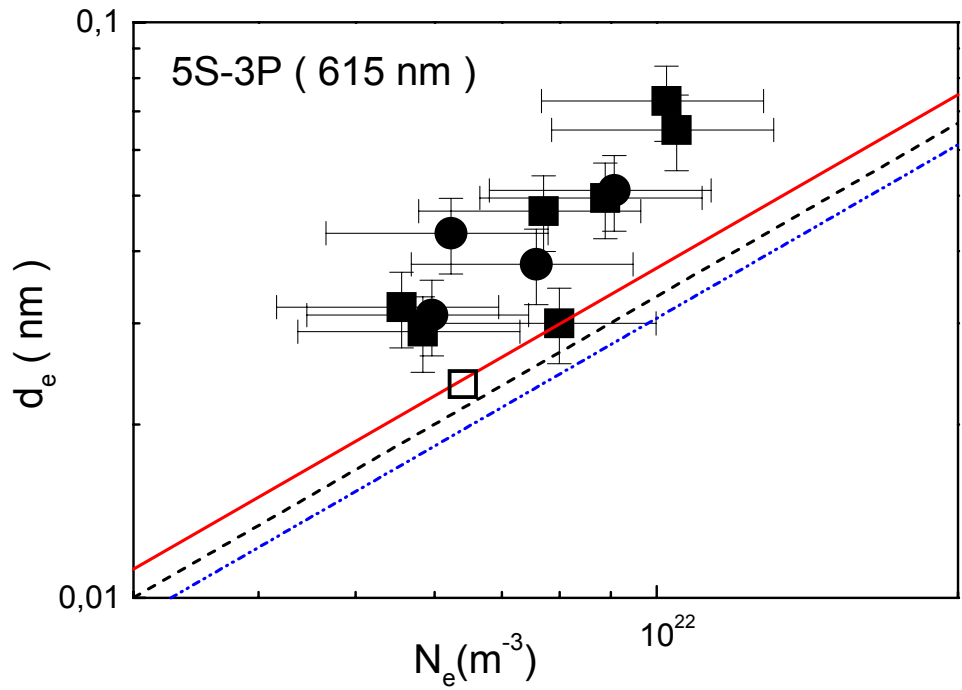

Fig.5c

**Fig. 6a** Results of the determination of the Stark-width for the 7S-3P transition. Explanation of the symbols: ■■■ this experiment, NaHg discharge, ●●● this experiment, NaCd discharge, – – – theory [9], —— theory [7]. Estimated experimental error is about ±25%. Claimed uncertainty of calculated width data is about 30% [7].

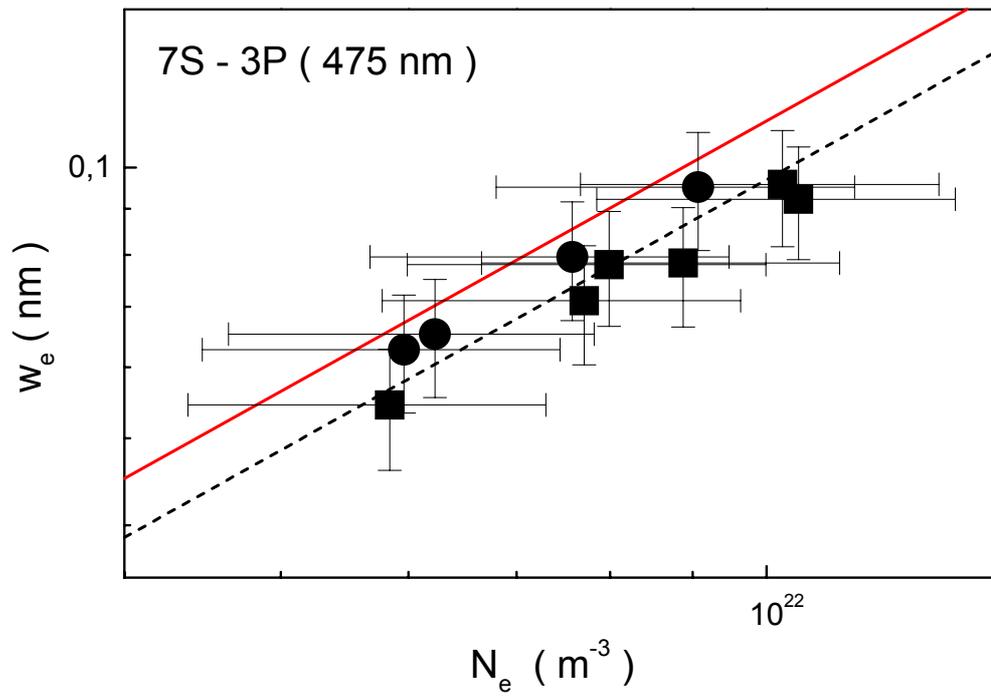

Fig. 6a

**Fig. 6b** Results of the determination of the Stark-width for the 6S-3P transition. Explanation of the symbols: ■■■ this experiment, NaHg discharge, ●●● this experiment, NaCd discharge, – – – theory [9], —— theory [7], –··–··· theory [8]. Note that the calculations of Griem [7] and Griem [8] practically coincide. Estimated experimental error is about ±25%. Claimed uncertainty of calculated shift data is about 30% [7].

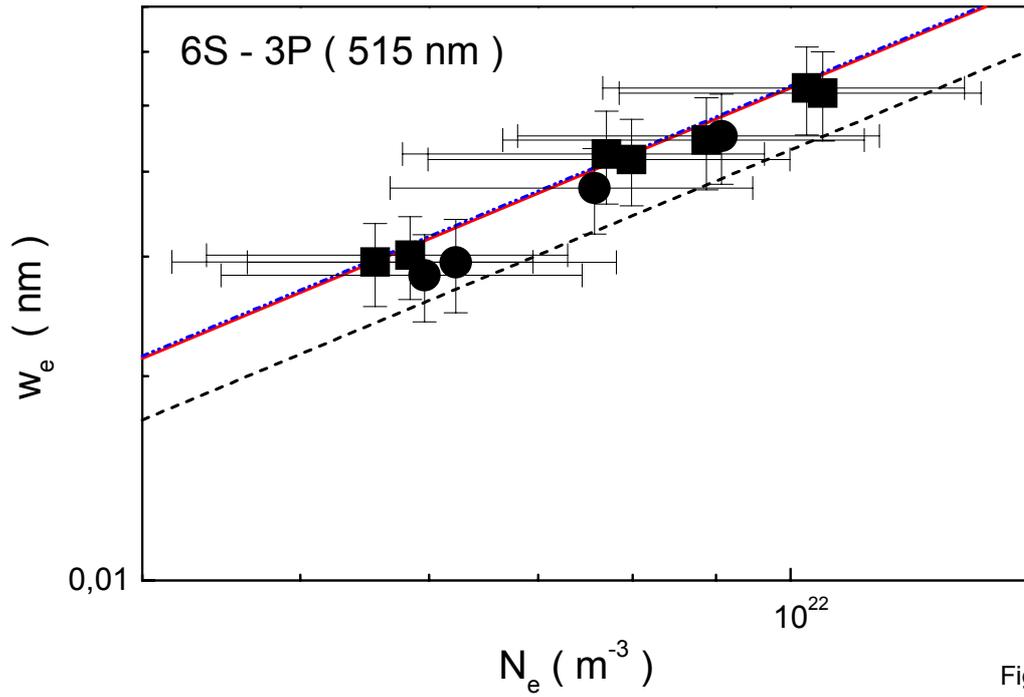

Fig. 6b

Miokovic et al

**Fig. 6c** Results of the determination of the Stark-width for the 5S-3P transition. Explanation of the symbols: ■■■ this experiment, NaHg discharge, ●●● this experiment, NaCd discharge, – – – theory [9], —— theory [7], –··–··· theory [8]. Estimated experimental error is about ±25%. Claimed uncertainty of calculated shift data is about 30% [7].

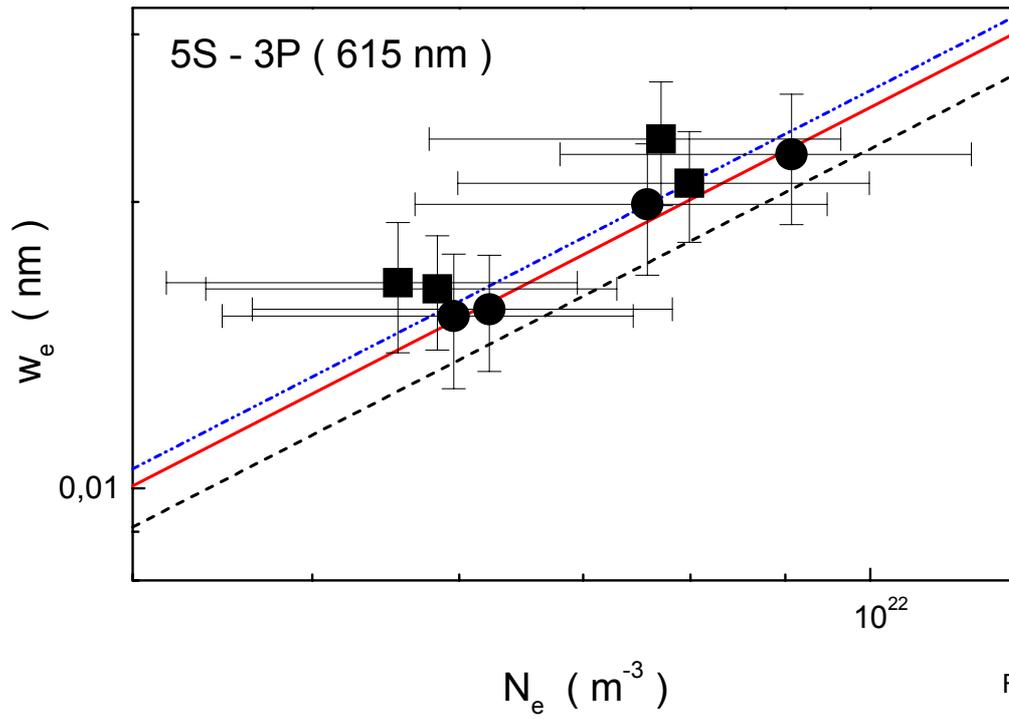

Fig. 6c

**Fig. 7a** Results of the measurements of the Stark-shift for the 6D-3P transition. Explanation of the symbols: ■■■ this experiment, NaHg discharge, ●●● this experiment, NaCd discharge, Δ Δ Δ experiment [11], – – – theory [9], —— theory [7], ······ RPA-theory [11]. Our estimated experimental error is ±25%. Claimed uncertainty of calculated shift data is about 30% [7]. Please, note the y-axis break at about 0.01 nm.

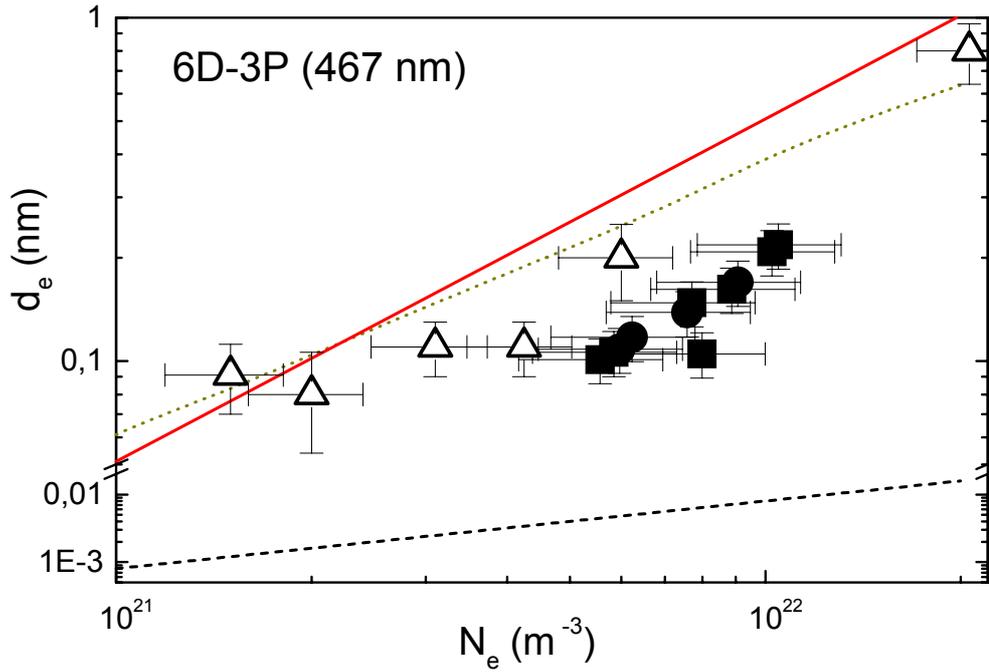

Fig. 7a

Miokovic et al

**Fig. 7b** Results of the measurements of the Stark-shift for the 5D-3P transition. Explanation of the symbols: ■■■ this experiment NaHg discharge, ●●● this experiment NaCd discharge, Δ Δ Δ experiment [11], ☐☐☐ experiment [10], – – – theory [9], —— theory [7], –··–··· theory [8], ······ RPA-theory [11]. Our estimated experimental error is ±25%. Claimed uncertainty of calculated shift data is about 30% [7].

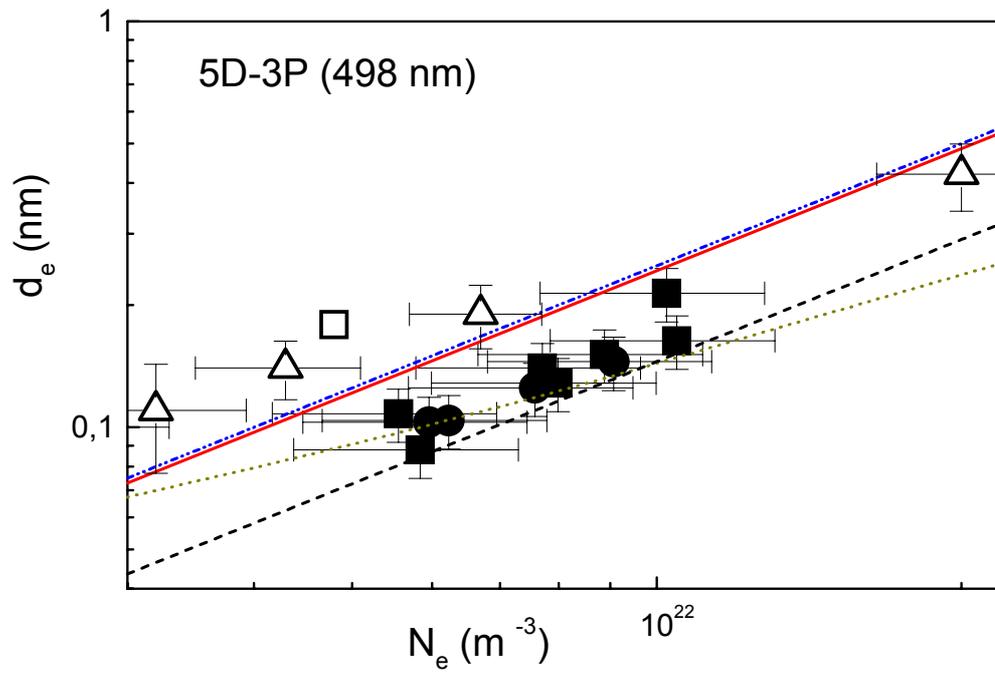

Fig. 7b

**Fig. 8** Measured and calculated Stark-shifts of sodium nS-3P lines (n=5,6,7) at $N_e=1*10^{16}$ cm$^{-3}$ plotted versus the upper state quantum number. Note a systematic disagreement of the experiment and calculations towards smaller quantum numbers (n=4, 5), and also substantial disagreement of the theoretical data in the range of higher quantum numbers, n ≥ 8. . Explanation of the symbols: ●●● this experiment , – – – theory [9], —— theory [7], –··–··· theory [8].

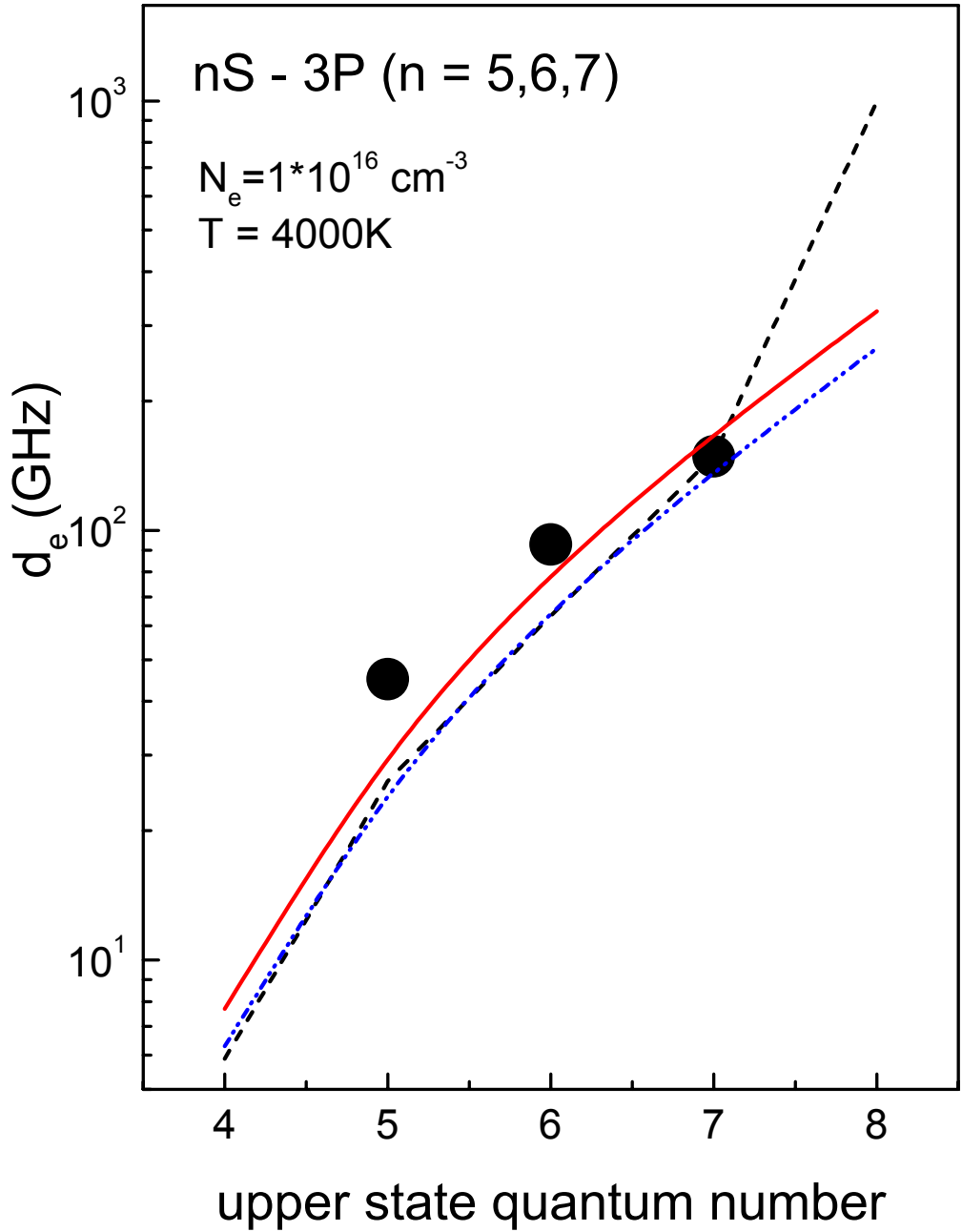

Fig. 8